\documentclass[a4paper,amsmath,amssymb]{jpconf}
\usepackage{graphicx}
\begin{document}
\title{Stochastic population oscillations in spatial predator-prey models}

\author{Uwe C. T\"auber}

\address{Department of Physics, Virginia Tech, Blacksburg, VA 24061-0435, USA}

\ead{tauber@vt.edu}

\begin{abstract}
It is well-established that including spatial structure and stochastic noise in
models for predator-prey interactions invalidates the classical deterministic 
Lotka--Volterra picture of neutral population cycles. 
In contrast, stochastic models yield long-lived, but ultimately decaying 
erratic population oscillations, which can be understood through a resonant 
amplification mechanism for density fluctuations. 
In Monte Carlo simulations of spatial stochastic predator-prey systems, one
observes striking complex spatio-temporal structures. 
These spreading activity fronts induce persistent correlations between 
predators and prey.
In the presence of local particle density restrictions (finite prey carrying 
capacity), there exists an extinction threshold for the predator population.
The accompanying continuous non-equilibrium phase transition is governed by the
directed-percolation universality class.
We employ field-theoretic methods based on the Doi--Peliti representation of 
the master equation for stochastic particle interaction models to (i) map the
ensuing action in the vicinity of the absorbing state phase transition to 
Reggeon field theory, and (ii) to quantitatively address fluctuation-induced 
renormalizations of the population oscillation frequency, damping, and 
diffusion coefficients in the species coexistence phase.
\end{abstract}

\section{Introduction}
\label{sec:Introduction}

Over the past decade, mathematical and computational tools from statistical 
physics have been increasingly and quite successfully applied to ecological 
problems, including attempts at a quantitative understanding of biodiversity 
\cite{May73}--\cite{Murray02}. % \cite{May73, Maynard74, Sigmund98, Murray02}.
In this context, physicists typically consider simplified idealized models that
hopefully capture the essential features of interacting biosystems; leaving 
aside some of the biological complexity allows the consistent incorporation of 
stochastic fluctuations and spatio-temporal correlations, whose crucial 
importance has long been recognized in the field \cite{Durrett99}, but is still
often neglected.

Predator-prey models defined via reaction-diffusion systems on a regular 
lattice, whose rate equations in the well-mixed mean-field limit reduce to the 
classic coupled Lotka--Volterra ordinary differential equations, constitute 
paradigmatic examples of the dynamics of two competing populations 
\cite{Matsuda92}--\cite{Boccara94}. % \cite{Matsuda92, Tome94, Boccara94}.
Monte Carlo simulations of these models, specifically in two dimensions, 
display a remarkable wealth of intriguing features (for a fairly recent 
overview, see, e.g., Ref.~\cite{Mobilia07}):
In contrast to the regular non-linear oscillations of the deterministic 
Lotka--Volterra model for which the population densities invariably return to 
their initial values (c.f. figure~\ref{fig:mfreq} below), computer simulations 
display persistent, but eventually decaying stochastic population oscillations 
(figure~\ref{fig:smosc}) \cite{Provata99}--\cite{Kowalik02}. % \cite{Provata99,
% Albano99, Lipowski99, Lipowska00, Monetti00, Droz01, Antal01, Kowalik02}.
In the absence of spatial degrees of freedom, these erratic population
oscillations may be understood through a resonant stochastic amplification 
mechanism \cite{McKane05} that drastically extends the transient time interval 
before any finite system ultimately reaches its absorbing stationary state, 
where the predator population becomes extinct \cite{Parker09}.
In spatially extended systems, it is well-known that the mean-field 
Lotka--Volterra reaction-diffusion equations allow for traveling wave solutions
\cite{Dunbar83}--\cite{Aguiar04}. % \cite{Dunbar83, Sherratt97, Aguiar04}.
In the corresponding stochastic spatial realizations, spreading activity fronts
(figure~\ref{fig:2dsnp}, \cite{Movies}) induce short-ranged but significant 
positive correlations of either species, and anti-correlations between the 
predator and prey populations, which have the effect of further enhancing the 
amplitude and life time of local population oscillations 
\cite{Mobilia07, Washenberger07}.
We have investigated various different variants of stochastic spatial 
Lotka--Volterra models for competing predator-prey populations, and found 
these intriguing spatio-temporal structures to be remarkably robust against
rather drastic changes of the detailed microscopic interaction rules 
\cite{Washenberger07, Mobilia06}, and even the introduction of quenched spatial
disorder in the reaction rates \cite{Dobramysl08}.

In this brief communication, I will provide an overview of our Monte Carlo
simulation results, specifically contrasting model variants with and without 
restrictions on the number of particles per lattice site. 
The former describe ecological systems with finite local carrying capacity, and
display a continuous non-equilibrium phase transition from an active species 
coexistence state to an absorbing phase wherein the predators become extinct.
Numerical evidence supports the general expectation 
\cite{Janssen81}--\cite{Janssen05}
% \cite{Janssen81, Grassberger82, Hinrichsen00, Janssen01, Odor04, Janssen05} 
that this extinction transition should be governed by the directed-percolation 
universality class \cite{Tome94, Boccara94}, \cite{Albano99}--\cite{Monetti00},
\cite{Antal01, Kowalik02}. % \cite{Tome94, Boccara94, Albano99, Lipowski99, 
% Lipowska00, Monetti00, Antal01, Kowalik02}.
I will then demonstrate how field-theoretic tools based on the Doi--Peliti 
representation of the master equation for stochastic interacting particle 
systems \cite{Doi76}--\cite{Peliti85} % \cite{Doi76, Grassberger80, Peliti85} 
(for recent reviews, see Refs.~\cite{Mattis98, Tauber05}), augmented with a 
means to incorporate restricted site occupation numbers \cite{Wijland01}, can 
be employed to gain a comprehensive understanding of fluctuation and 
correlation effects in Lotka--Volterra predator models.
Specifically, the effective action near the extinction transition in model 
variants with restricted site occupations will be explicitly mapped onto 
Reggeon field theory which describes the universal scaling of 
directed-percolation clusters \cite{Janssen81, Janssen05, Obukhov80, Cardy80}.
Moreover, expanding on the treatment in Ref.~\cite{Butler09}, I shall report a
computation of the fluctuation-induced renormalizations of the population 
oscillation frequency, damping, and diffusion coefficients in the species 
coexistence phase to lowest order in a perturbation expansion with respect to 
the predation rate \cite{Tauber11}.

\section{Stochastic lattice Lotka--Volterra models}

\subsection{Model variants and mean-field description}

We consider a two-species system of diffusing particles (with diffusion 
constant $D$) that undergo the following stochastic reactions:
\begin{eqnarray}
  &A \to \emptyset  \quad &{\rm with \ rate} \ \mu , \nonumber \\
  &A + B \to A + A  \quad &{\rm with \ rate} \ \lambda , 
\label{lvreac} \\
  &B \to B + B \quad &{\rm with \ rate} \ \sigma . \nonumber
\end{eqnarray}
The `predators' $A$ decay or die spontaneously at rate $\mu > 0$, whereas the 
`prey' $B$ produce offspring with rate $\sigma > 0$.
In the absence of the binary `predation' interaction with rate $\lambda$, the 
uncoupled first-order processes would naturally lead to predator extinction 
$a(t) = a(0) \, e^{- \mu t}$, and Malthusian prey population explosion
$b(t) = b(0) \, e^{\sigma t}$; here $a(t)$ and $b(t)$ respectively indicate the
$A$ / $B$ concentrations or population densities.
The binary predation reaction induces species coexistence through the 
non-linear interaction of both particle species.

In the simplest spatial realization of this stochastic reaction-diffusion 
model, both particle species are represented by unbiased random walkers on a
$d$-dimensional hypercubic lattice, and we allow an arbitrary number of 
particles per lattice site \cite{Washenberger07}.
All reactions (\ref{lvreac}) can then be implemented strictly on-site: 
Offspring particles are placed on the same lattice point as their parents, and 
the predation reaction happens only if an $A$ and a $B$ particle meet on the 
same lattice site.
If we then assume the populations to remain well mixed, and consequently ignore
both spatial fluctuations and correlations, we can approximately describe the
coupled reactions (\ref{lvreac}) through the associated mean-field rate
equations for spatially homogeneous concentrations 
$a(t) = \langle a({\vec x},t) \rangle$, $b(t) = \langle b({\vec x},t) \rangle$,
where $a({\vec x},t)$ and $b({\vec x},t)$ respectively denote the local 
predator and prey densities.
One then arrives at the classic Lotka--Volterra equations \cite{Murray02}, a
coupled set of two ordinary non-linear differential equations:
\begin{equation}
  \dot{a}(t) = \lambda \, a(t) \, b(t) - \mu \, a(t) \ , \quad
  \dot{b}(t) = \sigma \, b(t) - \lambda \, a(t) \, b(t) \ .
\label{lvreqa}
\end{equation}

\begin{figure}[ht]
\begin{center}
\includegraphics[width=7.8cm]{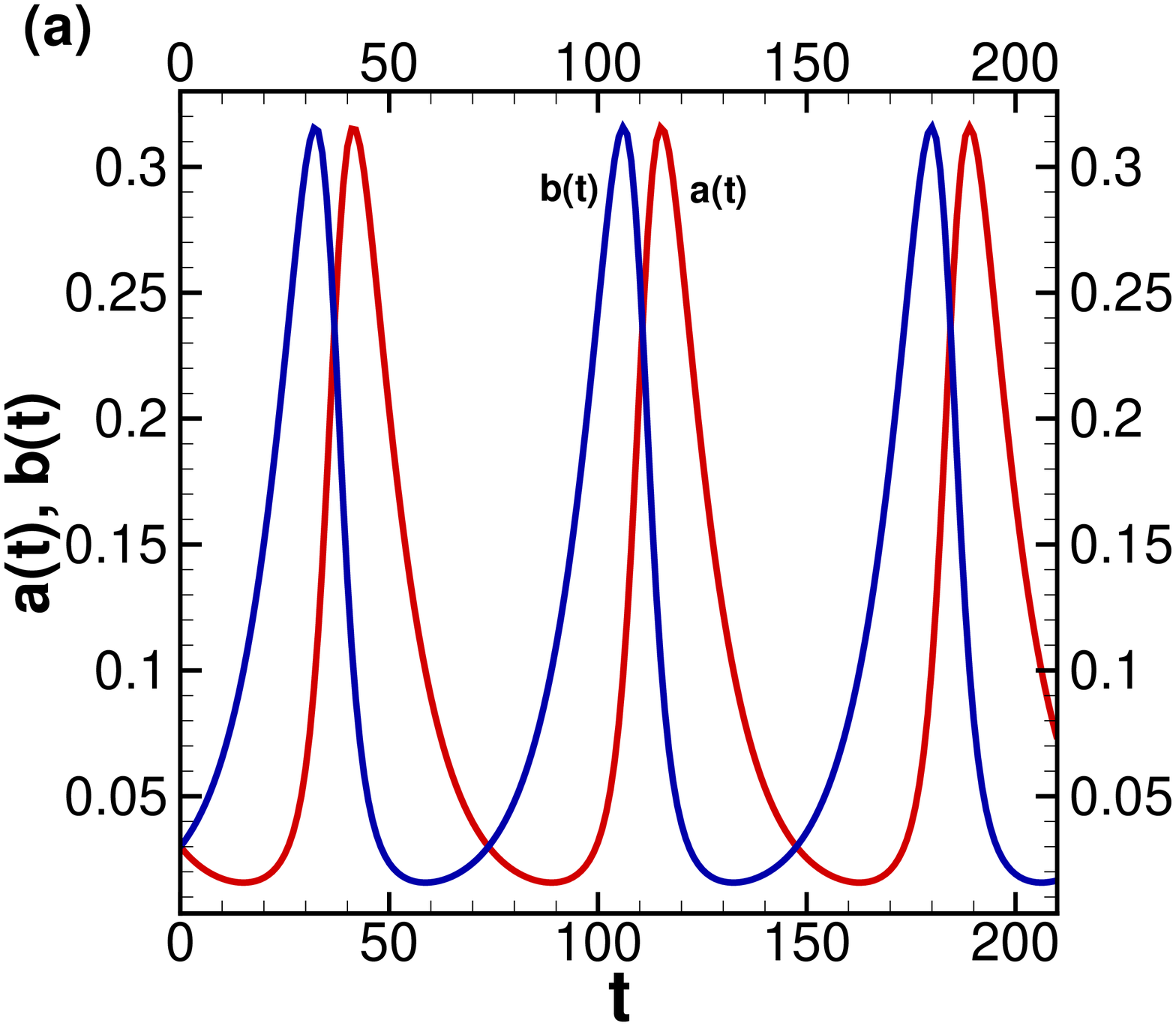} \ \
\includegraphics[width=7.8cm]{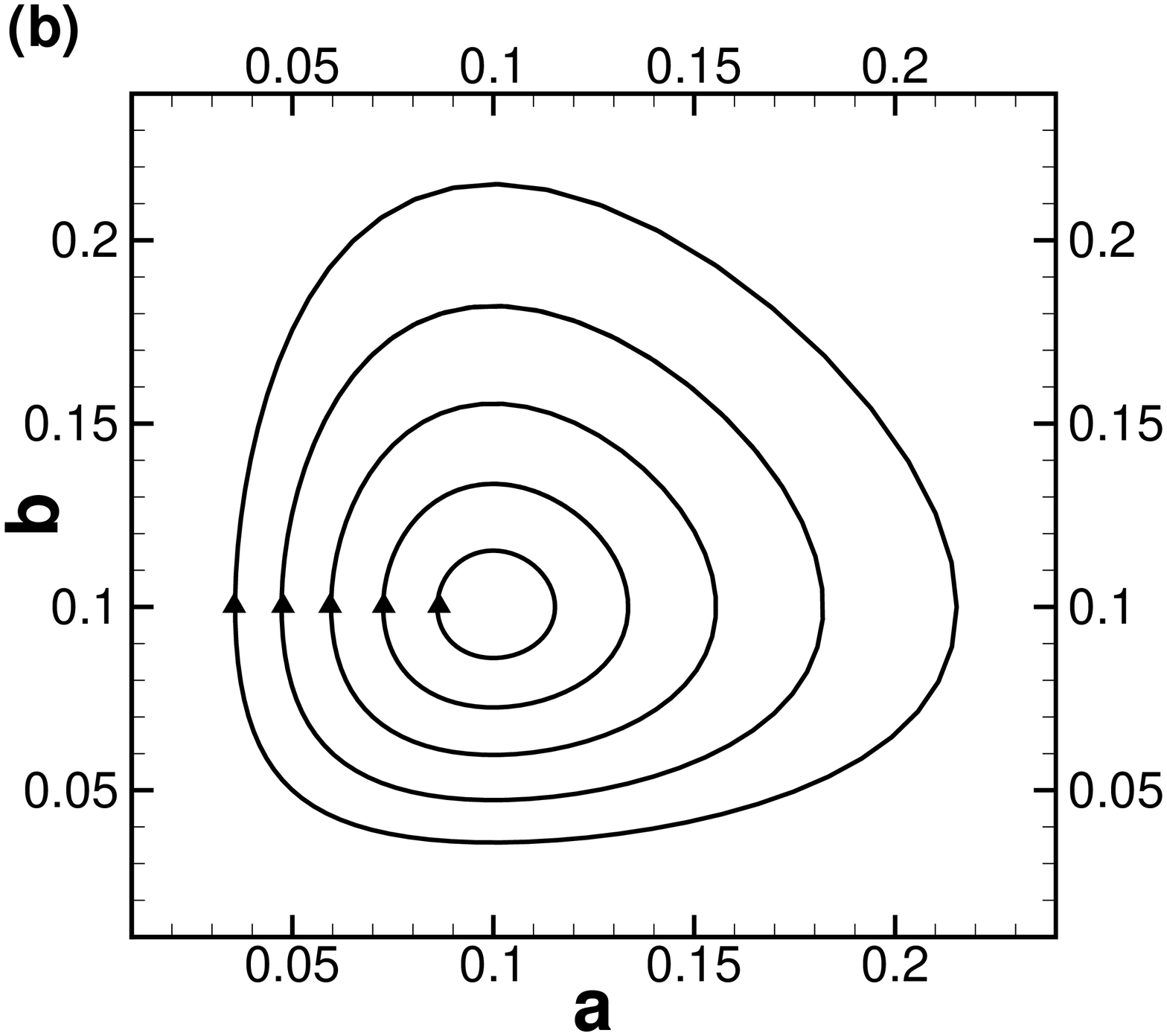}
\end{center}
\caption{\footnotesize Solutions of the coupled Lotka--Volterra mean-field rate
         equations (\ref{lvreqa}): 
	 (a) Non-linear predator $a(t)$ (red) and prey $b(t)$ (blue) population
         density oscillations; (b) periodic orbits in the $a$-$b$ phase plane.
	 For small amplitudes the oscillations become harmonic (circular 
         orbits) with frequency $\omega = \sqrt{\mu \, \sigma}$.
         (Reproduced with permission from Ref.~\cite{Washenberger07}, p.~4.)}
\label{fig:mfreq}
\end{figure}

The rate equations (\ref{lvreqa}) display three stationary states $(a_s,b_s)$,
namely the empty absorbing state with total population extinction $(0,0)$, 
which is obviously linearly unstable if $\sigma > 0$; a predator extinction 
absorbing state wherein the prey population diverges $(0,\infty)$, which for 
$\lambda > 0$ is also linearly unstable; and finally a species coexistence 
state $(a_u = \sigma/\lambda , b_u = \mu/\lambda)$, which however represents 
only a marginally stable fixed point with purely imaginary eigenvalues
$\pm i \sqrt{\mu \, \sigma}$ of the associated Jacobian stability matrix:
Linearizing eqs.~(\ref{lvreqa}) near $(a_u,b_u)$ results in the coupled
differential equations $\delta \dot{a}(t) = \sigma \, \delta b(t)$,
$\delta \dot{b}(t) = - \mu \, \delta a(t)$, which are readily solved by 
$\delta a(t) = \delta a(0) \, \cos\left( \sqrt{\mu \, \sigma} \, t \right) + 
\delta b(0) \, \sqrt{\sigma / \mu} \, \sin\left( \sqrt{\mu \, \sigma} \, t 
\right)$ and $\delta b(t) = - \delta a(0) \, \sqrt{\mu / \sigma} \, 
\sin\left( \sqrt{\mu \, \sigma} \, t \right) + \delta b(0) \, 
\cos\left( \sqrt{\mu \, \sigma} \, t \right)$, describing harmonic oscillations
about the center fixed point $(a_u,b_u)$ with frequency 
$\omega = 2 \pi \, f = \sqrt{\mu \, \sigma}$. 
Indeed, the phase space trajectories for the full non-linear coupled 
differential equations (\ref{lvreqa}) are determined by 
$d a / d b = a \, (\lambda \, b - \mu) / b \, (\sigma - \lambda \, a)$, with a 
conserved first integral 
\begin{equation}
  K(t) = \lambda [a(t) + b(t)] - \sigma \ln a(t) - \mu \ln b(t) = K(0) \ .
\label{mfcons}
\end{equation}
Consequently, as depicted in figure~\ref{fig:mfreq}, the solutions of the 
deterministic mean-field Lotka--Volterra model are closed orbits in phase 
space, i.e., regular periodic non-linear population oscillations whose 
amplitudes are fixed by the initial configuration.
Naturally, the precise periodic return to the initial concentration values does
not appear to be a very realistic feature.
In addition, the neutral cycles of the coupled mean-field rate equation system 
(\ref{lvreqa}) indicates that this deterministic mathematical model is 
fundamentally unstable with respect to slight modifications \cite{Murray02}.

One such modification that aims at rendering the Lotka--Volterra system more
relevant biologically is to introduce a finite carrying capacity (total 
particle density) $\rho > 0$ that limits the prey population growth, modeling,
e.g., the effect of limited food resources \cite{Murray02}.
Within the mean-field rate equation approximation, the second differential 
equation in (\ref{lvreqa}) is then replaced with
\begin{equation}
  \dot{b}(t) = 
  \sigma \, b(t) \left[ 1 - b(t) / \rho \right] - \lambda \, a(t) \, b(t) \ . 
\label{lvreqc}
\end{equation}
The non-trivial stationary states in this restricted Lotka--Volterra model are 
predator extinction and prey saturation $(0,\rho)$, linearly stable for 
$\lambda < \lambda_c = \mu / \rho$; and species coexistence $(a_r,b_r)$ with 
$b_r = \mu / \lambda$ and $a_r = (1 - \mu / \rho \lambda) \sigma / \lambda$, 
which both exists and is linearly stable provided the predation rate is
sufficiently large, $\lambda > \lambda_c$.
The eigenvalues of the Jacobian now acquire negative real parts,
$\epsilon_\pm = - \mu \, \sigma \left[ 1 \pm \sqrt{1 - 4 \rho \lambda \left( 
\rho \lambda / \mu - 1 \right) / \sigma} \right] / 2 \rho \lambda$, which 
implies an exponential approach to the stable fixed point $(a_r,b_r)$, 
replacing the neutral cycles of the unrestricted model (\ref{lvreqa}).
Moreover, for $\sigma > \sigma_s = 4 \lambda \, \rho \left( \rho \lambda / \mu 
- 1 \right) > 0$, or $\mu / \rho < \lambda < \lambda_s = \left( 1 + 
\sqrt{1 + \sigma / \mu} \right) \mu / 2 \rho$, the eigenvalues are real, 
indicating a nodal stable fixed point, whereas for $\sigma < \sigma_s$ or 
$\lambda > \lambda_s$, i.e., deep in the species coexistence phase, the
eigenvalues $\epsilon_\pm$ turn into a complex conjugate pair, and $(a_r,b_r)$ 
becomes a stable spiral singularity which is approached in a damped oscillatory
manner.
Adding spatial degrees of freedom, finite local carrying capacities can be
implemented in a lattice model through limiting the maximum occupation number 
per site for each species.
Most drastically, one can permit at most a single particle per lattice site 
\cite{Mobilia07}; the binary predation reaction then has to occur between 
predators and prey on adjacent nearest-neighbor sites, and new offspring needs 
to be placed on neighboring positions.
In that case, one can in fact entirely dispense with hopping processes, since 
all particle production reactions entail population spreading as well.
 
In summary, already within the mean-field rate approximation, a finite prey 
carrying capacity $\rho$, which can be viewed as the average result of local 
restrictions on the prey density originating from limited resources, crucially 
changes the phase diagram:
There emerges an extinction threshold (at $\lambda_c$ for fixed $\mu$) for the 
predator population, which in a spatially extended system becomes a genuine 
continuous active-to-absorbing non-equilibrium phase transition in the 
thermodynamic limit of infinite system size and time.

\subsection{Monte Carlo simulation results}

Various authors have studied stochastic lattice predator-prey models that in 
the well-mixed mean-field limit reduce to the classical Lotka--Volterra system
\cite{Matsuda92}--\cite{Boccara94}, \cite{Provata99}--\cite{Kowalik02}.
% \cite{Matsuda92, Tome94, Boccara94, Provata99, Albano99, Lipowski99, 
% Lipowska00, Monetti00, Droz01, Antal01, Kowalik02}.
In this section, I shall briefly discuss the pertinent results from our own
individual-based Monte Carlo simulation studies, performed mostly on 
two-dimensional square lattices with periodic boundary conditions. 
Technical details and more precise descriptions of the algorithms we have 
employed can be found in Refs.~\cite{Mobilia07} and \cite{Washenberger07}.

\begin{figure}[ht]
\begin{center}
\includegraphics[width=7.8cm]{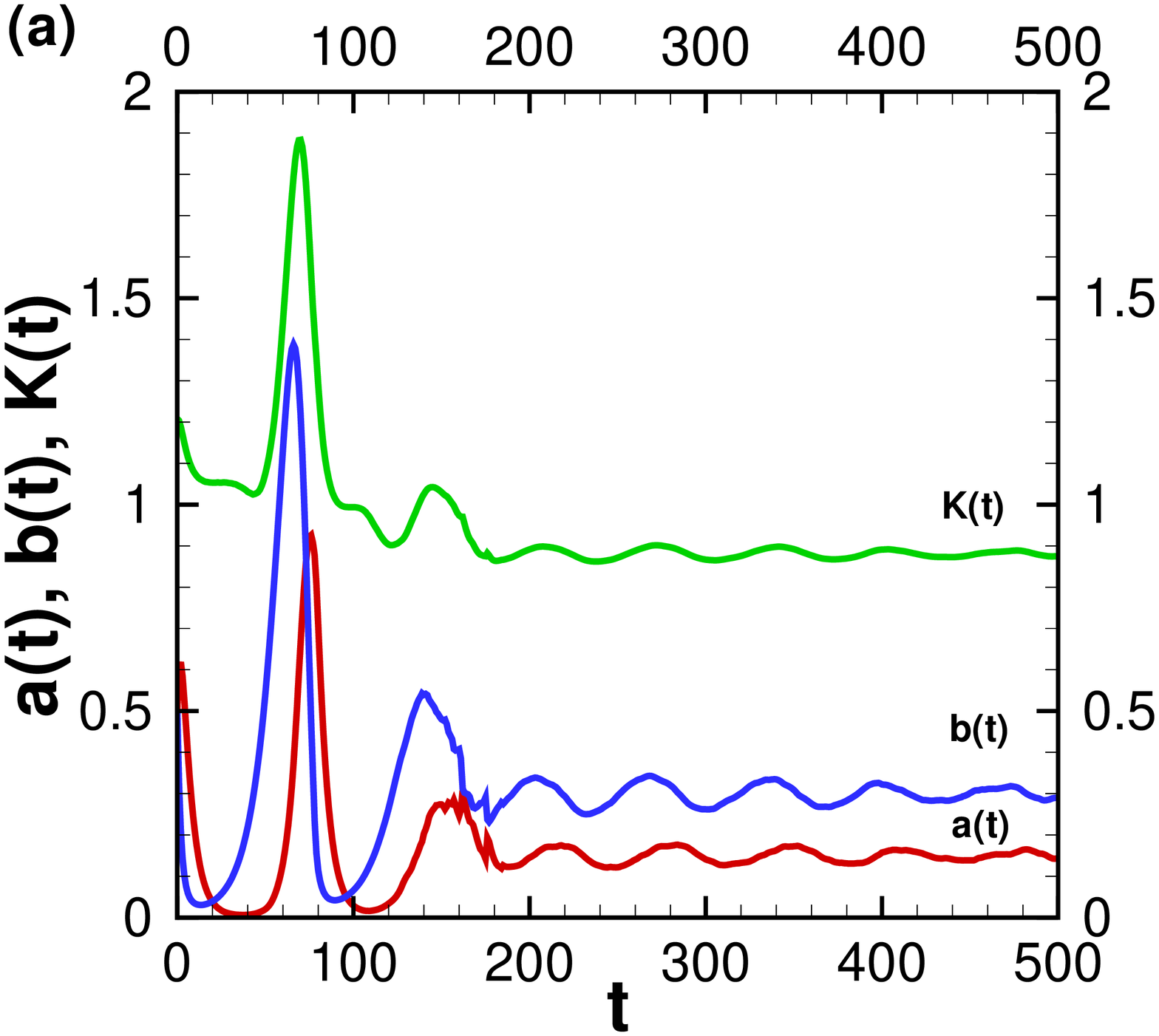} \ \
\includegraphics[width=7.8cm]{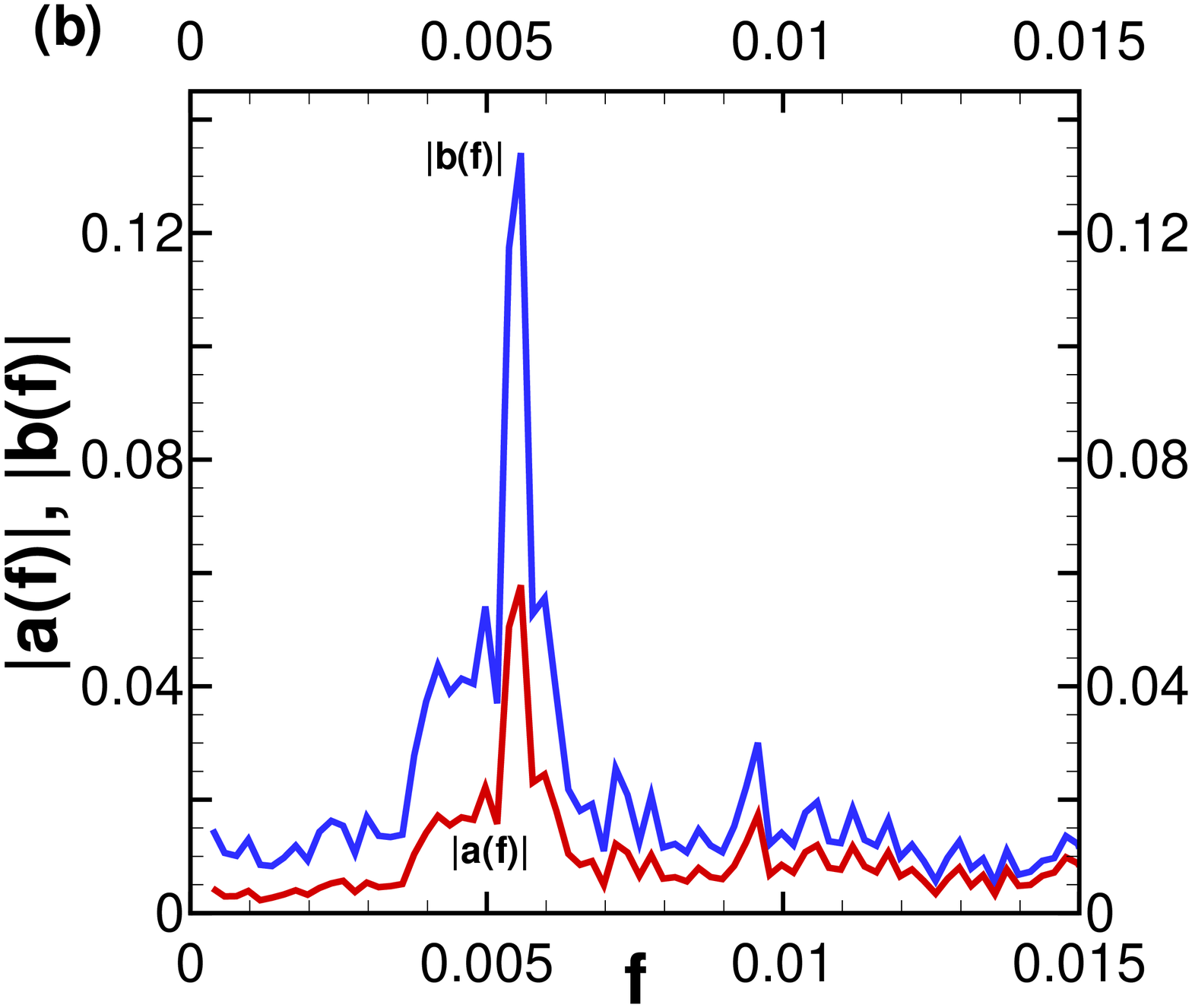}
\end{center}
\caption{\footnotesize Monte Carlo simulation data for a stochastic spatial 
         Lotka--Volterra system on a $1024 \times 1024$ square lattice with 
         periodic boundary conditions, in the absence of site occupation number
         restrictions: 
	 (a) Temporal evolution for the predator $a(t)$ (red) and prey $b(t)$ 
         (blue) population densities, and the quantity $K(t)$ (green) for
	 $\sigma = 0.1$, $\mu = 0.2,$, and $\lambda = 1.0$; 
	 (b) Fourier-transformed population density signals $|a(f)|$ and 
         $|b(f)|$ for $\sigma = 0.03$, $\mu = 0.1,$, and $\lambda = 1.0$. 
	 (Reproduced with permission from Ref.~\cite{Washenberger07}, pp.~9 and
         10.)}
\label{fig:smosc}
\end{figure}
\begin{figure}[ht]
\begin{center}
\includegraphics[width=7.65cm]{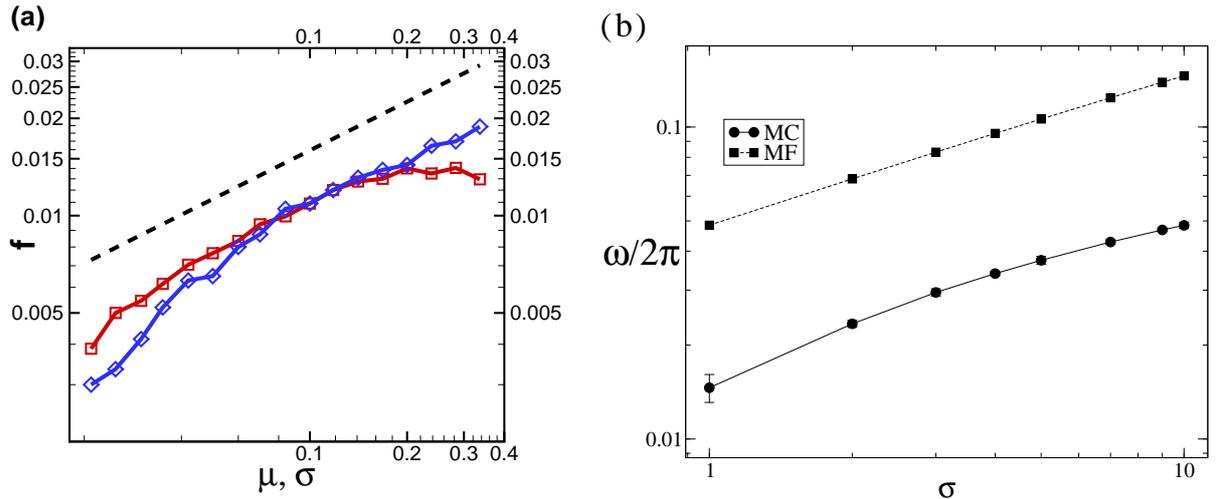} \ \
\includegraphics[width=8.0cm]{sgfreq.eps} 
\end{center}
\caption{\footnotesize Characteristic peak frequencies obtained form $|a(f)|$
         or $|b(f)|$, (a) as functions of the rates $\sigma$ (red squares) and
         $\mu$ (blue diamonds), with otherwise $\sigma = 0.1 = \mu$ and
         $\lambda = 1.0$ held fixed on a $1024 \time 1024$ lattice without site
         occupation restrictions;
         (b) as function of $\sigma$ with $\mu = 0.1$, $\lambda = 1.6$, $D = 0$
         on a $128 \times 128$ square lattice with at most a single particle 
         per lattice site (black line).
         The dashed black lines represent the oscillation frequency from 
	 linearized mean-field theory, $f = \sqrt{\mu \, \sigma}$.
	 (Reproduced with permission from Ref.~\cite{Washenberger07}, p.~10 and
         Ref.~\cite{Mobilia07}, p.~469.)}
\label{fig:renfr}
\end{figure}

Figure~\ref{fig:smosc} shows typical simulation data for the temporal evolution
of the total predator and prey particle densities in a two-dimensional 
stochastic lattice model with (almost) arbitrarily large site occupation
numbers and on-site reactions \cite{Washenberger07}.
One observes long-lived but clearly damped population oscillations that are 
actually quite independent of the initial state; neither are they caused by the
constancy of the first integral $K$, eq.~(\ref{mfcons}), that follows from the
deterministic rate equations:
As is apparent from the numerical data, in the stochastic spatial model $K(t)$ 
is manifestly time-dependent, and in fact traces the overall population 
oscillations.
We note that as the system size increases, the relative oscillation amplitudes 
become smaller; in the thermodynamic limit, the quasi-periodic population
fluctuations eventually die out entirely.
From the marked peaks in the Fourier-transformed concentration signals, 
$a(f) = \int a(t) \, e^{2 \pi i f t} \, dt$ for the predators, and similarly 
for the prey density, we may infer a characteristic oscillation frequency $f$.
As illustrated in figure~\ref{fig:renfr}, the typical population oscillation 
frequencies thus obtained roughly follow the square-root dependence on the 
rates $\mu$ and $\sigma$ as predicted by the linearized mean-field 
approximation, but with measurable deviations both for low and high rates. 
Yet the numerical frequency values are reduced by about a factor of four in the
stochastic spatially extended system, an apparent considerable downward 
renormalization caused by fluctuations and reaction-induced spatio-temporal 
correlations \cite{Washenberger07}.
Note also that figure~\ref{fig:renfr}(a) shows a remarkably similar functional 
dependence of $f$ on the rates $\mu$ and $\sigma$.

\begin{figure}[ht]
\begin{center}
\includegraphics[width=16cm]{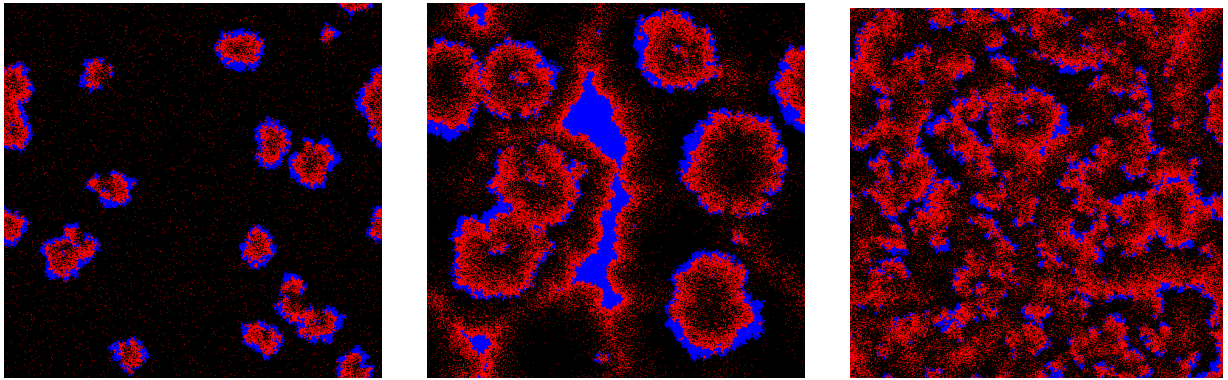}
\end{center}
\caption{\footnotesize Snapshots illustrating the evolution in time (from left 
         to right) of a two-dimensional stochastic lattice Lotka--Volterra 
         model (with $512 \times 512$ sites), incorporating local occupation 
	 number restrictions in the species coexistence phase, with rates 
	 $\sigma = 4.0$, $\mu = 0.1$, $\lambda = 2.2$, and $D = 0$, when the
         fixed point is a focus; the initial densities are $a(0) = 1/3 = b(0)$.
         The red, blue, and black dots respectively represent the prey, 
         predators, and empty lattice sites. 
	 (Reproduced with permission from Ref.~\cite{Mobilia07}, p.~464.)}
\label{fig:2dsnp}
\end{figure}

Very similar features are found in stochastic spatial Lotka--Volterra models
that incorporate stringent site occupation number restrictions (allowing only 
at most one particle on each site), deep in the species coexistence phase, 
i.e., for large predation rates, corresponding to a stable focal mean-field 
fixed point (stability matrix eigenvalues with negative real and non-vanishing 
imaginary parts). 
Both in the absence and presence of local density limitations, the coexistence
phase is governed by remarkably strong spatio-temporal fluctuations:
Striking spreading activity waves of prey closely followed by predators 
periodically sweep the system; any small surviving clusters of prey 
subsequently serve as sources for resurgent expanding prey-predator fronts 
\cite{Mobilia07}.
An average over these weakly coupled local oscillations then yields the total
population time traces depicted in figure~\ref{fig:smosc}.
These spreading activity fronts appear especially sharp for the site-restricted
model variants, as displayed in figure~\ref{fig:2dsnp}, whereas in realizations
with arbitrarily many particles per site, the fronts look more diffuse 
\cite{Movies}.
In either situation, one can employ stationary-state correlation functions to 
measure the spatial width $\sim 10 \ldots 20$ lattice sites of the spreading 
activity regions.
At roughly the same length scale, the cross-correlations of the $A$ and $B$ 
particles peak at a positive value before slowly decaying to zero; at shorter 
distances, the prey are naturally anti-correlated with the predators 
\cite{Mobilia07, Washenberger07}.

\begin{figure}[ht]
\begin{center}
\includegraphics[angle=-90,width=10.5cm]{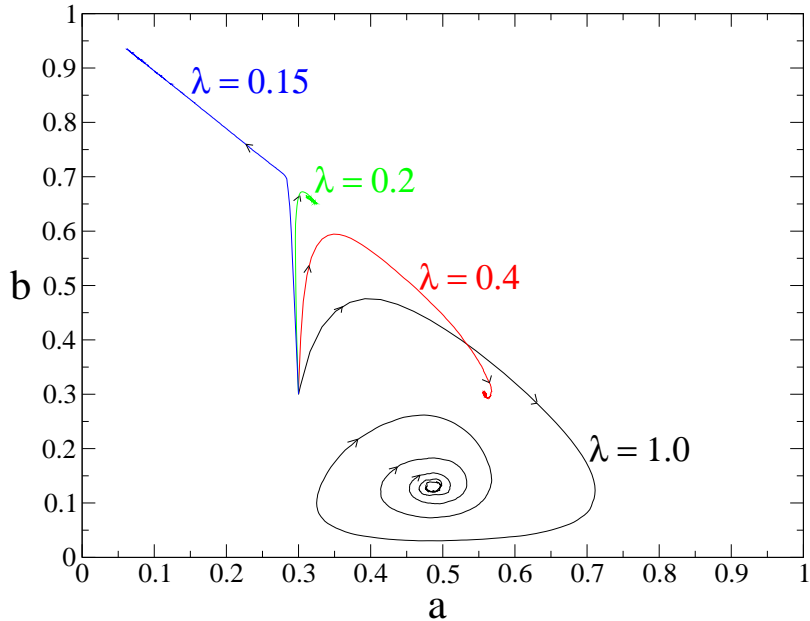}
\end{center}
\caption{\footnotesize Typical trajectories in the predator-prey phase space
         for a $512 \times 512$ stochastic lattice Lotka--Volterra system
         all initialized with $a(0) = 1/3 = b(0)$ and fixed rates 
         $\sigma = 4.0$, $\mu = 0.1$, and $D = 0$, but different predation 
         rates $\lambda = 0.15, 0.20, 0.40, 1.0$.
	 For small values of $\lambda$ (typically $\lambda < 0.4$) the fixed 
         point is a stable node, whereas for higher values of $\lambda$ one 
         observe the characteristic spirals that indicate a focus in phase 
         space.
	 (Reproduced with permission from Ref.~\cite{Mobilia07}, p.~463.)}
\label{fig:phpor}
\end{figure}
\begin{figure}[ht]
\begin{center}
\includegraphics[angle=-90,width=12cm]{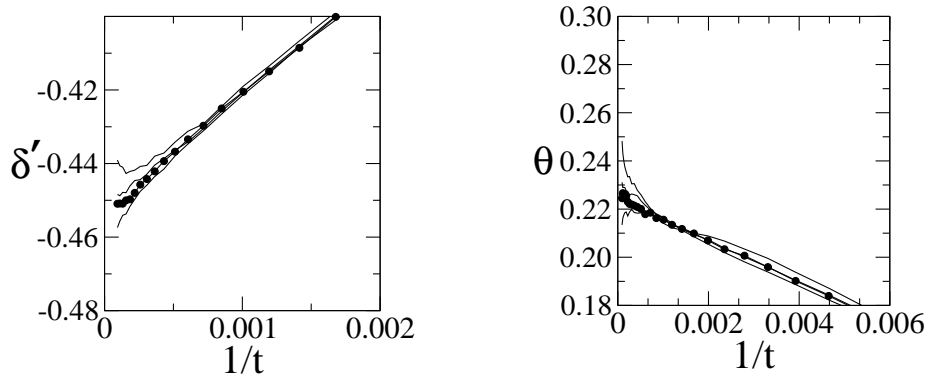}
\end{center}
\caption{\footnotesize Dynamical Monte Carlo simulation data to estimate the 
         critical point and scaling exponents for the predator extinction 
         threshold of the two-dimensional stochastic lattice Lotka--Volterra 
         model with site occupation restrictions (on a $512 \times 512$ 
         lattice):
	 The effective scaling exponents $- \delta'(t)$ vs. $1/t$ (left) and 
         $\theta(t)$ vs. $1/t$ (right) are depicted for four values of 
         $\lambda$ (from top to bottom): $0.1690$, $0.1689$, $0.1688$, and 
         $0.1687$, at fixed $\sigma = 4.0$, $\mu = 0.1$, and $D = 0$.
         (Reproduced with permission from Ref.~\cite{Mobilia07}, p.~470.)}
\label{fig:crtex}
\end{figure}

In the absence of spatial degrees of freedom, the observed persistent 
population oscillations can be mathematically understood by performing a 
systematic van-Kampen expansion about the absorbing steady state 
\cite{McKane05}.
The fluctuation corrections may then essentially be described through a damped
harmonic oscillator driven by white noise that will on occasion resonantly 
incite large-amplitude excursions away from the stable fixed point in the phase
plane.
In our spatial systems, we may also interpret the persistent population 
oscillations in the species coexistence regime through a similar mechanism, as
suggested by the bottom spiraling trajectory (for large predation rate 
$\lambda = 1.0$) in the phase portrait depicted in figure~\ref{fig:phpor}, 
which was obtained in simulation runs for stochastic lattice Lotka--Volterra 
models with restricted site occupancy \cite{Mobilia07}.
As the rate $\lambda$ is reduced (with all other parameters held constant) and 
the predators become less efficient, the stochastic lattice systems with site 
occupation restrictions qualitatively display the same scenarios as revealed by
the mean-field analysis for eqs.~(\ref{lvreqc}) with finite prey carrying 
capacity:
First, the focal stationary points in the phase plane are replaced by stable
nodes (real stability matrix eigenvalues); the population oscillations then 
cease, and no interesting spatial structures aside from localized activity 
clusters with meek fluctuations are seen (c.f. the trajectories for 
$\lambda = 0.4$ and $0.2$ in figure~\ref{fig:phpor}). 
At a sufficiently small critical value $\lambda_c$ ($\approx 0.1688$ here, see
figure~\ref{fig:crtex}), a predator extinction threshold is encountered, and
for $\lambda < \lambda_c$ ultimately the prey population fills the entire 
lattice.

For the predator population, the extinction threshold in stochastic spatial
Lotka--Volterra models with local particle density restrictions represents a
genuine continuous non-equilibrium phase transition in the thermodynamic and 
infinite-time limit.
Since no conserved quantities or disorder are present, one expects this active
to absorbing state phase transition to be described by the scaling exponents of
critical directed percolation \cite{Janssen81}--\cite{Janssen05}.
% \cite{Janssen81, Grassberger82, Hinrichsen00, Janssen01, Odor04, Janssen05} 
Heuristically, one may reason as follows:
The prey density is essentially uniform and constant $b \approx \rho$ near the 
critical point.
The Lotka--Volterra reactions (\ref{lvreac}) then basically reduce to 
$A \to \emptyset$ and $A \to A + A$; but since the $A$ population cannot 
multiply to arbitrarily large density (due to prey depletion), we need to add a
growth-limiting reaction such as $A + A \to A$, whereupon we arrive at the 
simplest microscopic reaction-diffusion model realization for the 
directed-percolation universality class 
\cite{Hinrichsen00, Odor04, Janssen05, Tauber05}.
This assertion is indeed supported by careful analysis of Monte Carlo 
simulation data \cite{Tome94}--\cite{Mobilia07}, 
\cite{Albano99}--\cite{Monetti00}, \cite{Antal01, Kowalik02}. 
% \cite{Tome94, Boccara94, Mobilia07, Albano99, Lipowski99, Lipowska00, 
% Monetti00, Antal01, Kowalik02}.
We performed dynamical Monte Carlo simulations starting from a single active
site with a predator particle in a lattice otherwise filled with prey, choosing
reaction rates in the vicinity of the extinction threshold.
The survival probability of predators at criticality is expected to decay
algebraically as $P(t) \sim t^{- \delta'}$, while the number of active sites
with predators should grow according to the power law $N(t) \sim t^\theta$
\cite{Hinrichsen00, Odor04, Janssen05}, with $\delta' \approx 0.451$ and 
$\theta \approx 0.230$ for directed percolation in two dimensions 
\cite{Hinrichsen00, Odor04}.
Figure~\ref{fig:crtex} shows the corresponding effective exponents as functions
of inverse time as measured in our Monte Carlo simulations for various values
of $\lambda$ with $\sigma = 4.0$, $\mu = 0.1$, and $D = 0$ held fixed
\cite{Mobilia07}.
From these data we infer $\lambda_c \approx 0.1688$ as best estimate for the
critical predation rate (compare with the mean-field prediction 
$\lambda_c = \mu = 0.1$), and the extrapolation to $t \to \infty$ yields very 
good agreement of the asymptotic critical exponents with the accepted 
directed-percolation values.

Simulations in one spatial dimension (on a circular domain) yield remarkable
differences between model variants with and without site occupation number
restrictions: 
In the former situation, the $A$ and $B$ particles quickly segregate into
distinct domains, with the predation reactions occurring only at the boundary.
The subsequent time evolution is governed by very slow coarsening induced by
merging predator domains \cite{Mobilia07}.
Without site occupation restrictions, in contrast we always observe the active
coexistence state \cite{Washenberger07}.
All the above statements of course pertain to sufficiently large lattices. 
In principle, any finite system with an absorbing steady state will eventually 
terminate in it; however, the associated survival times are expected to grow 
with system size according to a power law \cite{Parker09}, and for our lattices
are much longer than the duration of the simulations.

\section{Field-theoretic analysis}

\subsection{Field theory representation}

In the remainder of this paper, I shall describe how stochastic fluctuations,
internal reaction noise, and emerging correlations in spatial predator-prey
models can be systematically captured by means of a field-theoretic 
representation of the associated classical master equation, which is then 
amenable to analytic approximations.
Since the two-species Lotka--Volterra model (\ref{lvreac}) is defined via a 
diffusion-limited stochastic reaction system, we may employ the by now standard
Doi--Peliti framework to map the associated master equation onto a field theory
action \cite{Doi76}--\cite{Tauber05}.
% \cite{Doi76, Grassberger80, Peliti85, Mattis98, Tauber05}.
This approach is based on the fact that at any time the configurations in such 
systems are completely enumerated through specifying the occupation numbers of 
each species per lattice site, and that all occurring stochastic processes 
merely modify these local integer occupation numbers.
It is therefore natural to use bosonic creation and annihilation operators to
formally represent the system's temporal evolution which is given in terms of a
stochastic master equation.
Subsequently the continuum limit can be taken, which in the time domain is most
conveniently accomplished through a coherent-state path integral representation
for the evolution operator.  

For the diffusion-limited reactions (\ref{lvreac}) in $d$ spatial dimensions
one thus arrives at the following action \cite{Mobilia07} (see also 
Ref.~\cite{Butler09}) 
\begin{eqnarray}
  &&S[{\hat a},a;{\hat b},b] = \int \! d^dx \int \! dt \, \biggl[ \,
  {\hat a} \left( \partial_t - D_A \, \nabla^2 \right) a 
  + {\hat b} \left( \partial_t - D_B \, \nabla^2 \right) b \nonumber \\
  &&\quad\qquad\qquad\qquad\qquad\ + \mu \left( {\hat a} - 1 \right) a
  + \sigma \left( 1 - {\hat b} \right) {\hat b} \, b 
  + \lambda \left( {\hat b} - {\hat a} \right) {\hat a} \, a \, b \, \biggr] 
  \, ,
\label{lvact1}
\end{eqnarray}
with $e^{-S}$ providing the statistical weight for any observables that must be
functions of the local `density' fields $a({\vec x},t)$ and $b({\vec x},t)$.
The top line here obviously accounts for nearest-neighbor hopping processes 
in the continuum limit (through the inverse diffusion propagators with 
diffusivities $D_A$ and $D_B$ for the predators and prey, respectively).
The bottom line in (\ref{lvact1}) contains the stochastic reactions: 
spontaneous predator death with rate $\mu$, prey birth with rate $\sigma$, and 
predation with rate $\lambda$.
Note that each reaction process is represented by two contributions, 
originating from the gain and loss terms in the master equation.
One can easily reconstruct these contributions in the action by noting that the
second one directly reflects the reaction process itself through the 
annihilation operators $a$, $b$ and creation operators ${\hat a}$, ${\hat b}$,
whereas the first one encodes the `order' of the corresponding reaction (i.e.,
which powers of the concentrations ${\hat a} \, a$ and ${\hat b} \, b$ enter 
the rate equations).
It is important to realize that the Doi--Peliti action faithfully contains all
stochastic fluctuations associated with the underlying microscopic processes,
namely discrete finite-number fluctuations and internal reaction noise
\cite{Tauber05}.
Following van~Wijland's analysis \cite{Wijland01}, restricted site occupation
numbers or finite local carrying capacities $\rho$ for the prey species can be 
incorporated in this bosonic formalism through the replacement
$\sigma \to \sigma \, e^{- {\hat b} \, b / \rho}$ in the $B$ particle 
reproduction term.
(Alternatively, a growth-limiting reaction such as $B + B \to B$ could have 
been added.)

The associated classical field equations follow from the stationarity 
conditions $\delta S / \delta a = 0 = \delta S / \delta b$, always solved by
${\hat a} = 1 = {\hat b}$ (actually just reflecting probability conservation 
\cite{Tauber05}), and $\delta S / \delta {\hat a}({\vec x},t) = 0 = 
\delta S / \delta {\hat b}({\vec x},t)$, which yields precisely the mean-field 
rate equations augmented by diffusion terms.
Indeed, for $\rho = \infty$ one arrives at eqs.~(\ref{lvreqa}), while expanding
to first order in $\rho^{-1}$ recovers eq.~(\ref{lvreqc}).
It is then convenient to perform a field shift according to 
${\hat a} = 1 + {\tilde a}$, ${\hat b} = 1 + {\tilde b}$, whereupon the action
becomes, again to lowest order in the inverse carrying capacity,
\begin{eqnarray}
  &&S[{\tilde a},a;{\tilde b},b] = \int \! d^dx \int \! dt \, \biggl[ \, 
  {\tilde a} \left( \partial_t - D_A \, \nabla^2 + \mu \right) a 
  + {\tilde b} \left( \partial_t - D_B \, \nabla^2 - \sigma \right) b
  \nonumber \\
  &&\quad\qquad\qquad\qquad\qquad\ - \sigma \, {\tilde b}^2 \, b 
  + \sigma \, \rho^{-1} \, (1 + {\tilde b})^2 \, {\tilde b} \, b^2 
  - \lambda \, (1 + {\tilde a}) \, ({\tilde a} - {\tilde b}) \, a \, b \,
  \biggr] \, . 
\label{lvact2} 
\end{eqnarray}
In the following, the `microscopic' field theory action (\ref{lvact2}) will 
serve as the starting point (i) for further manipulations to identify the 
universality class of the continuous active to absorbing state phase transition
at the predator extinction threshold, and (ii) to compute the 
fluctuation-induced renormalization to lowest order in the predation rate for
the population oscillation frequency and damping, as well as the diffusion 
coefficient in the two-species coexistence phase.

\subsection{Extinction transition and directed percolation}

Our goal is to construct an effective field theory \cite{Mobilia07} that 
describes the universal scaling properties near the non-equilibrium phase 
transition at $\lambda_c \approx \mu / \rho$ where the predators go extinct, 
and the prey fill the entire lattice: $a_s = 0$, $b_s \approx \rho$.
Consequently we transform the action (\ref{lvact2}) to new fluctuating fields 
$c = b_s - b$ with $\langle c \rangle = 0$, and ${\tilde c} = - {\tilde b}$:
\begin{eqnarray}
  &&S[{\tilde a},a;{\tilde c},c] = \int \! d^dx \int \! dt \, \biggl[ \, 
  {\tilde a} \left( \partial_t - D_A \, \nabla^2 + \mu - \lambda \, b_s \right)
  a + {\tilde c} \left( \partial_t - D_B \nabla^2 
  + (2 b_s / \rho - 1) \, \sigma \right) c \nonumber \\
  &&\quad\qquad\qquad\qquad\qquad\ + \sigma \, b_s (2 b_s / \rho - 1) \, 
  {\tilde c}^2 - \sigma \, \rho^{-1} \, b_s^2 \, {\tilde c}^3 - \sigma \, 
  (4 b_s / \rho - 1) \, {\tilde c}^2 \, c \nonumber \\
  &&\quad\qquad\qquad\qquad\qquad\ - \sigma \, \rho^{-1} \, (1 + {\tilde c}^2)
  \, {\tilde c} \, c^2 + 2 \sigma \, \rho^{-1} \, {\tilde c}^2 \, (c + b_s \, 
  {\tilde c}) \, c \nonumber \\
  &&\quad\qquad\qquad\qquad\qquad\ - \lambda \, b_s \left( {\tilde a}^2 + 
  (1 + {\tilde a}) \, {\tilde c} \right) a + \lambda \left( 1 + {\tilde a} 
  \right) \left( {\tilde a} + {\tilde c} \right) a \, c \, \biggr] \, .
\label{dpact1}
\end{eqnarray}
Next we note that the birth rate is a relevant parameter in the renormalization
group sense, which scales to infinity under scale transformations; this
observation simply expresses the fact that fluctuations of the nearly uniform 
prey population become strongly suppressed through the `mass' term 
$\propto \sigma$ for the $c$ fields.
It is therefore appropriate to introduce rescaled fields 
$\phi = \sqrt{\sigma} \, c$ and ${\tilde \phi} = \sqrt{\sigma} \, {\tilde c}$, 
and subsequently take the limit $\sigma \to \infty$, which yields the 
drastically reduced effective action
\begin{equation}
  S_\infty[{\tilde a},a;{\tilde \phi},\phi] = \int \! d^dx \int \! dt \, 
  \biggl[ \, {\tilde a} \left( \partial_t - D_A \, \nabla^2 + \mu 
  - \lambda \, b_s \right) a - \lambda \, b_s \, {\tilde a}^2 \, a 
  + {\tilde \phi} \, \phi + b_s \,  {\tilde \phi}^2 \, \biggr] \, .
\label{dpact2}
\end{equation}

As a final step, one needs to add a growth-limiting process for the predator
population, for example through the binary coagulation reaction 
$A + A \to A$ with rate $\tau$.
Since the fields $\phi$ and ${\tilde \phi}$ only appear as a bilinear form in
the action (\ref{dpact2}), they can readily be integrated out, leaving 
\begin{equation}
  S_\infty[{\widetilde {\psi}},\psi] = \int \! d^dx \int \! dt \, \Biggl[ \, 
  {\widetilde \psi} \biggl( \partial_t + D_A \, (r_A - \nabla^2) \biggr) \psi  
  - u \, {\widetilde \psi} \left( {\widetilde \psi} - \psi \right) \psi 
  + \tau \, {\widetilde \psi }^2 \, \psi^2 \, \biggr] \, ,
\label{dprft}
\end{equation}
where $\psi = a \, \sqrt{\lambda \, b_s / \tau}$, 
${\widetilde \psi} = {\tilde a} \, \sqrt{\tau / \lambda \, b_s}$, 
$r_A = (\mu - \lambda \, b_s) / D_A$, and $u = \sqrt{\tau \, \lambda \, b_s}$.
This new effective non-linear coupling $u$ becomes dimensionless at $d_c = 4$, 
signifying the upper critical dimension for this field theory. 
Near four dimensions, the quartic term $\propto \tau$ constitutes an irrelevant
contribution in the renormalization group sense and may be omitted for the
analysis of universal asymptotic power laws at the phase transition.
The action (\ref{dprft}) then becomes identical to Reggeon field theory, which
is known to describe the critical scaling exponents for directed percolation
\cite{Janssen81, Janssen05, Obukhov80, Cardy80}.
This mapping to Reggeon field theory \cite{Mobilia07} firmly corroborates the 
expectation that the predator extinction threshold is governed by the 
directed-percolation universality class \cite{Tome94, Boccara94}, 
\cite{Albano99}--\cite{Monetti00}, \cite{Antal01, Kowalik02}, % \cite{Tome94, 
% Boccara94, Albano99, Lipowski99, Lipowska00, Monetti00, Antal01, Kowalik02},
which features quite prominently in phase transitions to absorbing states 
\cite{Janssen81, Grassberger82}, even in multi-species systems 
\cite{Janssen01}. 
The universal scaling properties of critical directed percolation are 
well-understood and quantitatively characterized to remarkable accuracy, both 
numerically through extensive Monte Carlo simulations and analytically by means
of renormalization group calculations (for overviews, see 
Refs.~\cite{Hinrichsen00, Odor04, Janssen05}).

\subsection{Fluctuation corrections in the coexistence phase}

In order to address fluctuation corrections in the predator-prey coexistence
phase \cite{Tauber11}, we start again from the Doi--Peliti field theory action 
(\ref{lvact2}), and introduce the proper fluctuating fields 
$c = a - \langle a \rangle$ and $d = b - \langle b \rangle$:
\begin{equation} 
  a = \frac{\sigma}{\lambda} \left( 1 - \frac{\mu}{\rho \, \lambda} 
  + A_c \right) + c \ , \quad 
  b =  \frac{\mu}{\lambda} \left( 1 + B_c \right) + d \ .
\label{cphcd}
\end{equation}
Here, the mean-field values for the stationary densities have been taken into
account already, such that the counter-terms $A_c$ and $B_c$, which are 
naturally determined by the conditions 
$\langle c \rangle = 0 = \langle d \rangle$, contain only fluctuation 
contributions.
The bilinear terms in the ensuing action may then readily be diagonalized by 
introducing new fields $\varphi_\pm$ and ${\widetilde \varphi}_\pm$,
\begin{eqnarray}
   &&c = \frac{1}{\sqrt{2 \mu}} \, \left[ \, \varphi_+ + \varphi_-
   - \frac{\gamma_0}{i \omega_0} \, (\varphi_+ - \varphi_-) \right] \, , \quad 
   d = \sqrt{\frac{\mu}{2}} \ \frac{\varphi_+ - \varphi_-}{i \omega_0} 
   \nonumber \\ 
   &&{\tilde a} = \sqrt{\frac{\mu}{2}} \ \frac{{\widetilde \varphi}_+ 
   - {\widetilde \varphi}_-}{i \omega_0} \, , \quad
   {\tilde b} = \frac{1}{\sqrt{2 \mu}} \, \left[ \, {\widetilde \varphi}_+ + 
   {\widetilde \varphi}_- + \frac{\gamma_0}{i \omega_0} \, 
   ({\widetilde \varphi}_+ - {\widetilde \varphi}_-) \right] \, ,
\label{cphph}
\end{eqnarray}
with the mean-field (or `bare') oscillation frequency and damping constant
(see also Ref.~\cite{Butler09})
\begin{equation}
   \omega_0^2 = \mu \, \sigma \left( 1 - \frac{\mu}{\rho \, \lambda} \right) 
   - \gamma_0^2 \ , \quad 
   \gamma_0 = \frac{\sigma \, \mu}{2 \, \rho \, \lambda} \ .
\label{bfrdm}
\end{equation}
Note that $\omega_0^2 = \mu \, \sigma$ and $\gamma_0 \to 0$ as 
$\rho \to \infty$: 
There is no damping of the mean-field oscillations in the absence of local
carrying capacity restrictions.

In the following, we shall consider equal diffusivities $D_A = D_0 = D_B$; the
harmonic propagators in the diagonalized theory then read in Fourier space
\begin{equation}
  \langle {\widetilde \phi}_\pm({\vec q},\omega) \, \phi_\pm({\vec q}',\omega')
  \rangle_0 = 
  \frac{\pm i \omega_0}{- i \omega + D_0 \, q^2 \pm i \omega_0 + \gamma_0} \ 
  (2 \pi)^{d+1} \, \delta({\vec q} + {\vec q}') \, \delta(\omega + \omega') \ .
\label{props}
\end{equation}
Along with two two-point noise sources and several non-linear vertices, these
propagators form the building blocks for the Feynman diagrams that graphically
represent the different contributions in a perturbation expansion in terms of
the non-linear coupling $\lambda$ \cite{Tauber11}.
To lowest non-trivial (`one-loop') order, only the noise and three-point
vertices are needed to determine the counter-terms $A_c$ and $B_c$, as well as
to compute the fluctuation corrections to the bare propagators (\ref{props}). 
From the ensuing one-loop expressions, one may infer renormalized versions of 
the diffusivity $D_R$, oscillation frequency $\omega_R$, and damping 
$\gamma_R$.
In addition, one finds that in the absence of site occupation restrictions 
(i.e., for infinite local prey carrying capacity $\rho$), the stochastic 
spatial fluctuations generate a damping term, just as seen in the lattice 
simulations.
These perturbational calculations are fairly straightforward, but lengthy and 
somewhat tedious; details will be reported elsewhere \cite{Tauber11}.
Here I merely provide the explicit results for the renormalized parameters in
several space dimensions.

For $d = 1$ and $d = 2$, the expressions for the renormalized oscillation 
frequency become singular in the limit $\gamma_0 \to 0$; in the list below, 
only the leading terms in $\gamma_0$ are retained:
\begin{eqnarray}
  &d = 1: \
  &D_R = D_0 + \frac{3 \, \lambda}{64\, \sqrt{2}} \ \sqrt{\frac{D_0}{\omega_0}}
  \ \left[ 1 + \frac{1}{12} \left( \sqrt{\frac{\sigma}{\mu}} - 
  \sqrt{\frac{\mu}{\sigma}} \right) + \frac{3}{4} \left( \frac{\sigma}{\mu} + 
  \frac{\mu}{\sigma} \right) \right] + {\cal O}(\lambda^2) \ , \nonumber \\
  &&\gamma_R = \frac{\lambda}{8 \, \sqrt{2}} \, \sqrt{\frac{\omega_0}{D_0}} \ 
  \left[ 1 + \frac{3}{4} \left( \sqrt{\frac{\sigma}{\mu}} - 
  \sqrt{\frac{\mu}{\sigma}} \right) - \frac{3}{4} \left( \frac{\sigma}{\mu} + 
  \frac{\mu}{\sigma} \right) \right] + {\cal O}(\lambda^2) \ , \nonumber \\
  &&\omega_R = \omega_0 - \frac{\lambda}{16} \, 
  \frac{\mu \, \sigma}{\sqrt{D_0 \, \gamma_0}} \ \left[ 1 + \frac{1}{2} \left( 
  \frac{\sigma}{\mu} + \frac{\mu}{\sigma} \right) \right] \nonumber \\
  &&\qquad + \frac{\lambda}{8 \, \sqrt{2}} \, \sqrt{\frac{\omega_0}{D_0}} \ 
  \left[ 1 - \frac{57}{32} \, \sqrt{\frac{\sigma}{\mu}} + \frac{25}{32} \,
  \sqrt{\frac{\mu}{\sigma}} + \frac{1}{32} \left( \frac{\sigma}{\mu} + 
  \frac{\mu}{\sigma} \right) \right] + {\cal O}(\lambda^2) \ . 
\label{renfr1} \\
  &d = 2: \
  &D_R = D_0 + \frac{\lambda}{96 \, \pi} \left[ 1 + 2 \left( \frac{\sigma}{\mu}
  + \frac{\mu}{\sigma} \right) \right] + {\cal O}(\lambda^2) \ , \nonumber \\
  &&\gamma_R = \frac{\lambda}{64} \, \frac{\omega_0}{D_0} \left[ \frac{6}{\pi} 
  \left( \sqrt{\frac{\sigma}{\mu}} - \sqrt{\frac{\mu}{\sigma}} \right) 
  - \left( \frac{\sigma}{\mu} + \frac{\mu}{\sigma} \right) \right] 
  + {\cal O}(\lambda^2) \ , \nonumber \\
  &&\omega_R = \omega_0 - \frac{\lambda}{32 \, \pi} \, \frac{\omega_0}{D_0} \, 
  \ln\frac{\omega_0}{\gamma_0} \ \left[ 1 + \frac{1}{2} \left( 
  \frac{\sigma}{\mu} + \frac{\mu}{\sigma} \right) \right] \nonumber \\
  &&\qquad + \frac{3 \, \lambda}{32 \, \pi} \, \frac{\omega_0}{D_0} \, 
  \left[ 1 - \frac{\pi}{3} \, \sqrt{\frac{\sigma}{\mu}} - \frac{1}{4} \left( 
  \frac{\sigma}{\mu} + \frac{\mu}{\sigma} \right) \right] + {\cal O}(\lambda^2)
  \ .
\label{renfr2}
\end{eqnarray}
Notice that the infrared singularities encountered in the limit 
$\gamma_0 \to 0$ cancel for the renormalized diffusivity $D_R$ and the
fluctuation-generated damping $\gamma_R$.
In dimensions $d < 2$, the leading fluctuation correction to the oscillation
frequency diverges as $(\omega_0 / \gamma_0)^{1 - d/2}$, acquiring a 
logarithmic dependence in two dimensions; it is negative, and symmetric under 
formal rate exchange $\mu \leftrightarrow \sigma$ (c.f. the top lines in the 
above one-loop results for $\omega_R$).
If we interpret $\gamma_0$ in the above equations as a small, self-consistently
determined damping, these features are in remarkable agreement with our earlier
Monte Carlo observations displayed in figure~\ref{fig:renfr}:
Fluctuations and correlations induced by the stochastic reaction processes 
induce a strong downward numerical renormalization of the oscillation 
frequency, with very similar functional dependence on the rates $\mu$ and 
$\sigma$.

In three dimensions, we may set the bare damping constant to zero (or 
$\rho \to \infty$) to obtain
\begin{eqnarray}
  &d = 3: \
  &D_R = D_0 - \frac{\lambda}{384\, \sqrt{2}\,\pi}\ \sqrt{\frac{\omega_0}{D_0}}
  \ \left[ 1 + \frac{9}{4} \left( \sqrt{\frac{\sigma}{\mu}} - 
  \sqrt{\frac{\mu}{\sigma}} \right) - \frac{13}{4} \left( \frac{\sigma}{\mu} + 
  \frac{\mu}{\sigma} \right) \right] + {\cal O}(\lambda^2) \ , \nonumber \\
  &&\gamma_R = \frac{\lambda}{16 \, \sqrt{2}\, \pi} \left( \frac{\omega_0}{D_0}
  \right)^{3/2} \left[ - 1 + \frac{3}{4} \left( \sqrt{\frac{\sigma}{\mu}} - 
  \sqrt{\frac{\mu}{\sigma}} \right) - \frac{1}{4} \left( \frac{\sigma}{\mu} + 
  \frac{\mu}{\sigma} \right) \right] + {\cal O}(\lambda^2) \ , 
\label{renfr3} \\
  &&\omega_R = \omega_0 + \frac{\lambda}{128 \, \sqrt{2} \, \pi} \left( 
  \frac{\omega_0}{D_0} \right)^{3/2} \left[ 1 - \frac{13}{4} \, 
  \sqrt{\frac{\sigma}{\mu}} - \frac{19}{4} \, \sqrt{\frac{\mu}{\sigma}} - 
  \frac{13}{4} \left( \frac{\sigma}{\mu} + \frac{\mu}{\sigma} \right) \right]
  + {\cal O}(\lambda^2) \ . \nonumber
\end{eqnarray}
In higher dimensions $d \geq 4$, the fluctuation corrections become formally
ultraviolet-divergent, and thus a finite cut-off $\Lambda$ in momentum space 
must be implemented; e.g., in four dimensions one finds
\begin{eqnarray}
  &d = 4: \,
  &D_R = D_0 - \frac{\lambda}{512 \, \pi} \, \frac{\omega_0}{D_0}
  \left[ 1 + \frac{1}{\pi} \left( \sqrt{\frac{\sigma}{\mu}} - 
  \sqrt{\frac{\mu}{\sigma}} \right) \ln\!\left(\! 1 + 
  \frac{\Lambda^4}{\omega_0^2/D_0^2} \right) - \left( \frac{\sigma}{\mu} + 
  \frac{\mu}{\sigma} \right) \right] + {\cal O}(\lambda^2) \ , \nonumber \\
  &&\gamma_R = \frac{\lambda}{32\, \pi^2} \left( \frac{\omega_0}{D_0} \right)^2
  \left[ 1 - \frac{1}{2}\, \ln\!\left(\! 1 + \frac{\Lambda^4}{\omega_0^2/D_0^2}
  \right) + \frac{3 \, \pi}{8} \left( \sqrt{\frac{\sigma}{\mu}} - 
  \sqrt{\frac{\mu}{\sigma}} \right) - \frac{1}{4} \left( \frac{\sigma}{\mu} + 
  \frac{\mu}{\sigma} \right) \right] \nonumber \\
  &&\qquad + {\cal O}(\lambda^2) \ , \nonumber \\
  &&\omega_R = \omega_0 + \frac{\lambda}{256 \, \pi} \left( 
  \frac{\omega_0}{D_0} \right)^2 \biggl[ 1 - \frac{2}{\pi} \, 
  \sqrt{\frac{\mu}{\sigma}} \, \ln\!\left(\! 1 + 
  \frac{\Lambda^4}{\omega_0^2/D_0^2} \right) - \frac{5}{2 \, \pi^2} \left( 
  \sqrt{\frac{\sigma}{\mu}} - \sqrt{\frac{\mu}{\sigma}} \right) \nonumber \\
  &&\qquad\qquad\qquad\qquad\qquad - \left( \frac{\sigma}{\mu} + 
  \frac{\mu}{\sigma} \right) \biggr] + {\cal O}(\lambda^2) \ .
\label{renfr4} 
\end{eqnarray}
We finally remark that the effective expansion parameter in this fluctuation
perturbation series in $d$ dimensions is 
$(\lambda / \omega_0) \, (\omega_0 / D_0)^{d/2}$.

\section{Concluding remarks}

In conclusion, in this contribution I have reviewed the most striking features
of stochastic predator-prey models on regular lattices that in the well-mixed
mean-field limit reduce to the celebrated Lotka--Volterra model.
It turns out that the spatially extended stochastic systems display both richer
behavior than the associated deterministic rate equations, and are actually 
also more robust with respect to modifications of model and algorithmic 
details:
Spatial predator-prey systems in the species coexistence phase are generically 
characterized by the emergence of persistent spatio-temporal structures, 
namely continually expanding and merging activity fronts, leading to transient 
oscillations for the total (or mean) particle densities.
Fluctuations in the two-species coexistence phase are remarkably and unusually
strong; they markedly alter the oscillation frequency as compared to the
(linearized) mean-field prediction, and in addition generate damping.
Restricting the (local) prey population through a growth-limiting finite
carrying capacity induces a genuine continuous non-equilibrium extinction phase
transition for the predators.
I have also outlined how the Doi--Peliti field theory representation of the
associated master equation can be employed to (i) demonstrate that this active
to absorbing state transition is governed by the universal scaling exponents of
critical directed percolation, and (ii) permits a systematic perturbational 
approach to compute the fluctuation-induced renormalizations of the population 
spreading and oscillation parameters in the coexistence phase.

It remains to be elucidated which of the standard mathematical models in 
ecology, population dynamics, and chemical kinetics, many of which are 
frequently just discussed on the level of mean-field rate equations, are 
similarly strongly affected by stochastic fluctuations and intrinsic 
correlations.
Perhaps unexpectedly, stochastic spatial variants of cyclic three-species 
predator-prey systems that are often referred to as rock-paper-scissors models 
represent an intriguing counter-example:
Lattice simulations of these reaction-diffusion systems hardly show any 
noticeable fluctuation effects, both for model variants with conserved and 
non-conserved total particle number, despite the formation of striking spiral
structures in the latter, so-called May--Leonard model, see 
Refs.~\cite{He10, He11} (and further references therein).

\ack
The author warmly thanks the organizers of CMDS-12 for their kind invitation to
participate at this very stimulating conference.
Fruitful collaborations and insightful discussions with Ulrich Dobramysl,
Erwin Frey, Ivan Georgiev, Qian He, Swapnil Jawkar, Rahul Kulkarni, 
Gabriel Martinez, Mauro Mobilia, Tim Newman, Michel Pleimling, 
Beate Schmittmann, Siddharth Venkat, Mark Washenberger, and Royce Zia are 
gratefully acknowledged.


\section*{References}
\begin{thebibliography}{99}

\bibitem{May73}
  May R M 1973
  {\em Stability and complexity in model ecosystems},
  (Princeton: Princeton University Press)

\bibitem{Maynard74}
  Maynard Smith J 1974 
  {\em Models in ecology} 
  (Cambridge: Cambridge University Press)

\bibitem{Sigmund98}  
  Hofbauer J and Sigmund K 1998	
  {\em Evolutionary games and population dynamics} 
  (Cambridge: Cambridge University Press)

\bibitem{Murray02}
  Murray J D 2002
  {\em Mathematical biology}, Vols. I and II
  (New York: Springer, 3rd ed.)

\bibitem{Durrett99}
  Durrett R 1999
  {\it SIAM Review} {\bf 41} 677 

\bibitem{Matsuda92}
  Matsuda H, Ogita N, Sasaki A and Sat$\rm \overline{o}$ K 1992
  {\it Prog. Theor. Phys.} {\bf 88} 1035

\bibitem{Tome94}
  Satulovsky J E and Tom\'e T 1994
  {\it Phys. Rev. E} {\bf 49} 5073

\bibitem{Boccara94}
  Boccara N, Roblin O and Roger M 1994
  {\it Phys. Rev. E} {\bf 50} 4531

\bibitem{Mobilia07}
  Mobilia M, Georgiev I T and T\"auber UC 2007 
  {\it J. Stat. Phys.} {\bf 128} 447
  [doi: 10.1007/s10955-006-9146-3]

\bibitem{Provata99}
  Provata A, Nicolis G and Baras F 1999
  {\it J. Chem. Phys.} {\bf 110} 8361

\bibitem{Albano99}
  Rozenfeld A F and Albano E V 1999
  {\it Physica A} {\bf 266} 322

\bibitem{Lipowski99}
  Lipowski A  1999 
  {\it Phys. Rev. E} {\bf 60} 5179

\bibitem{Lipowska00}
  Lipowski A and Lipowska D 2000
  {\it Physica A} {\bf 276} 456

\bibitem{Monetti00}
  Monetti R, Rozenfeld A F and Albano E V 2000
  {\it Physica A} {\bf 283} 52

\bibitem{Droz01}
  Droz M and P\c ekalski A 2001
  {\it Phys. Rev. E} {\bf 63} 051909

\bibitem{Antal01}
  Antal T and Droz M 2001
  {\it Phys. Rev. E} {\bf 63} 056119

\bibitem{Kowalik02}
  Kowalik M, Lipowski A and Ferreira A L 2002
  {\it Phys. Rev. E} {\bf 66} 066107

\bibitem{McKane05}
  McKane A J and Newman T J 2005 
  {\it Phys. Rev. Lett.} {\bf 94} 218102

\bibitem{Parker09}
  Parker M and Kamenev A 2009
  {\it Phys. Rev. E} {\bf 80} 021129

\bibitem{Dunbar83}
  Dunbar S R 1983
  {\it J. Math. Biol.} {\bf 17} 11

\bibitem{Sherratt97}
  Sherratt J, Eagen B T and Lewis M A 1997 
  {\it Phil. Trans. R. Soc. Lond. B} {\bf 352} 21

\bibitem{Aguiar04}  
  de Aguiar M A M, Rauch A M and Bar-Yam Y 2004
  {\it J. Stat. Phys.} {\bf 114} 1417

\bibitem{Movies}
  Monte Carlo simulation movies are available at 
  {\tt http://www.phys.vt.edu/$\,\sim$tauber/PredatorPrey/movies/}

\bibitem{Washenberger07}
  Washenberger M J, Mobilia M and T\"auber UC 2007
  {\it J. Phys. Condens. Matter} {\bf 19} 065139
  [doi: 10.1088/0953-8984/19/6/065139]

\bibitem{Mobilia06} 
  Mobilia M, Georgiev I T and T\"auber U C 2006 
  {\it Phys. Rev. E} {\bf 73}, 040903(R)

\bibitem{Dobramysl08}
  Dobramysl U and T\"auber U C 2008
  {\it Phys. Rev. Lett.} {\bf 101}, 258102

\bibitem{Janssen81}
  Janssen H K 1981
  {\it Z. Phys. B} {\bf 42} 151

\bibitem{Grassberger82}
  Grassberger P 1982
  {\it Z. Phys. B} {\bf 47} 365

\bibitem{Hinrichsen00}
  Hinrichsen H 2000
  {\it Adv. Phys.} {\bf 49} 815

\bibitem{Janssen01} 
  Janssen H K 2001
  {\it J. Stat. Phys.} {\bf 103} 801

\bibitem{Odor04}
  \'Odor G 2004
  {\it Rev. Mod. Phys.} {\bf 76} 663 

\bibitem{Janssen05}
  Janssen H K and T\"auber U C 2005
  {\it Ann. Phys.} {\bf 315} 147

\bibitem{Doi76}
  Doi M 1976
  {\it J. Phys. A: Math. Gen.} {\bf 9} 1465

\bibitem{Grassberger80}
  Grassberger P and Scheunert P 1980
  {\it Fortschr. Phys.} {\bf 28} 547

\bibitem{Peliti85}
  Peliti L 1985
  {\it J. Phys. (France)} {\bf 46} 1469; 1479

\bibitem{Mattis98}
  Mattis D C and Glasser M L 1998
  {\it Rev. Mod. Phys.} {\bf 70} 979

\bibitem{Tauber05}
  T\"auber U C, Howard M and Vollmayr-Lee B P 2005
  {\it J. Phys. A: Math. Gen.} {\bf 38} R79

\bibitem{Wijland01}
  van Wijland F 2001
  {\it Phys. Rev. E} {\bf 63} 022101

\bibitem{Obukhov80}
  Obukhov S P 1980
  {\it Physica A} {\bf 101} 145

\bibitem{Cardy80}
  Cardy J L and Sugar R L 1980
  {\it J. Phys. A: Math. Gen.} {\bf 13} L423

\bibitem{Butler09}
  Butler T and Reynolds D 2009
  {\it Phys. Rev. E} {\bf 79} 032901

\bibitem{Tauber11}
  T\"auber U C 2011 
  {\it manuscript in preparation}

\bibitem{He10}
  He Q, Mobilia M and T\"auber U C 2010
  {\it Phys. Rev. E} {\bf 82} 051909

\bibitem{He11}
  He Q, Mobilia M and T\"auber U C 2011
  {\it Eur. Phys. J. B} in press

\end{thebibliography}
\end{document}